# Ultrafast photothermoelectric effect in Dirac semimetallic $Cd_3As_2$ revealed by terahertz emission

Wei Lu[1,2], Zipu Fan[2], Yunkun Yang[3], Junchao Ma[2], Jiawei Lai[2], Xiaoming Song[1], Xiao Zhuo[2], Zhaoran Xu[2], Jing Liu[1], Xiaodong Hu[1], Shuyun Zhou[4,5], Faxian Xiu[3], Jinluo Cheng[6] & Dong Sun[2,5,*]

[1]State Key Laboratory of Precision Measurement Technology and Instruments, School of Precision Instruments and Opto-electronics Engineering, Tianjin University, Tianjin 300072, China

[2]International Center for Quantum Materials, School of Physics, Peking University, Beijing 100871, China

[3]State Key Laboratory of Surface Physics and Department of Physics, Fudan University, Shanghai 200433, China

[4]State Key Laboratory of Low Dimensional Quantum Physics and Department of Physics, Tsinghua University, Beijing, 100084, China

[5]Collaborative Innovation Center of Quantum Matter, Beijing 100871, China

[6]Changchun Institute of Optics, Fine Mechanics and Physics, Chinese Academy of Sciences, Changchun 130033, China

*Corresponding author. Email: sundong@pku.edu.cn (D.S.)

**Abstract**: The thermoelectric effects of topological semimetals have attracted tremendous research interest because many topological semimetals are excellent thermoelectric materials and thermoelectricity serves as one of their most important potential applications. In this work, we reveal the transient photothermoelectric response of Dirac semimetallic $Cd_3As_2$, namely the photo-Seebeck effect and photo-Nernst effect, by studying the terahertz (THz) emission from the transient photocurrent induced by these effects. Our excitation polarization and power dependence confirm that the observed THz emission is due to photothermoelectric effect instead of other nonlinear optical effect. Furthermore, when a weak magnetic field (~0.4 T) is applied, the response clearly indicates an order of magnitude enhancement on transient photothermoelectric current generation compared to the photo-Seebeck effect. Such enhancement supports an ambipolar transport nature of the photo-Nernst current generation in $Cd_3As_2$. These results highlight the enhancement of thermoelectric performance can be achieved in topological Dirac semimetals based on the Nernst effect, and our transient studies pave the way for thermoelectric devices applicable for high field circumstance when nonequilibrium state matters. The large THz emission due to highly efficient photothermoelectric conversion is comparable to conventional semiconductors through optical rectification and photo-Dember effect.

Thermoelectric (TE) properties, which determine heat-to-electricity energy conversion of materials, lie at the center toward TE applications. Traditionally, semimetals are usually regarded as bad TE materials, because they naturally possess two types of carriers which counterbalance each other's contribution to the induced TE voltage, leading to a reduced thermopower compare to semiconductors which usually has unipolar transports[1]. Interestingly, recent TE transport measurements on Weyl semimetals have shown that both types of carriers can contribute to the transverse TE voltage constructively through Nernst effect when a magnetic field is applied perpendicular to the applied temperature gradient, because electrons and holes are forced to deflect in opposite directions by the magnetic force[2-7]. Taking advantage of Nernst effect, four times higher thermopower factor compared to conventional Seebeck response is achieved on NbP[4], which is comparable to that of the state-of-the-art thermoelectric materials (see Supplementary Table 1 for a summary of thermoelectric coefficients of typical materials).

Different from steady state TE response, which is usually slow because it involves the heat transport dynamics of lattice, the photothermoelectric (PTE) response upon photoexcitation can be ultrafast as it is dominated by the transient carrier temperature, which is induced by the highly nonequilibrium carriers excited by ultrafast laser pulses[8-12]. The instantaneous transient states after the photoexcitation can support transient ambipolar transport and provide ambipolar TE effect. As such response is usually on the picosecond timescale in semimetals[9, 12-14], such transient current is difficult to be measured by conventional transport measurement. Instead, the picosecond transient current emits in terahertz (THz) frequency range, thus can be characterized by measuring its THz emission[15-17].

The ultrafast PTE response is an important aspect of photoexcited carrier dynamics, which is crucial for high field/speed electronic and optoelectronic device applications. With ultrahigh electron mobility and ultrafast carrier dynamics, topological semimetals are suitable for such applications[11, 13, 18, 19]. In these devices, the carriers are accelerated by the electric field which is similar to the photoexcited state after photoexcitation[9, 14, 20-24]. It is equally important for optoelectronic related applications, such as light harvesting and detection, where light interaction with materials is involved and the simultaneous transient optical process involves the TE response of the materials[8, 10, 11, 25-27].

In this work, we reveal the transient ambipolar TE current response of typical Dirac semimetal $Cd_3As_2$ after ultrafast photoexcitation under magnetic field. The transient thermal gradient is created by the thickness gradient of the as-grown $Cd_3As_2$ sample after ultrafast photoexcitation. As a contactless approach, the THz emission from the transient TE current is studied at room temperature. Our results clearly distinguish the transverse Seebeck response and the Nernst response from $Cd_3As_2$ after photoexcitation. According to the THz emission amplitudes, the Nernst effect is one order of magnitude larger than the Seebeck effect with a relatively weak magnetic field $B \sim 0.4$ T, and the large enhancement is attributed to the constructive ambipolar transport nature under magnetic field. Furthermore, the efficiency of THz emission from $Cd_3As_2$ is comparable to that of typical semiconductor THz sources. This work may open opportunities for high speed TE devices, taking advantage of their PTE effects.

**Results**

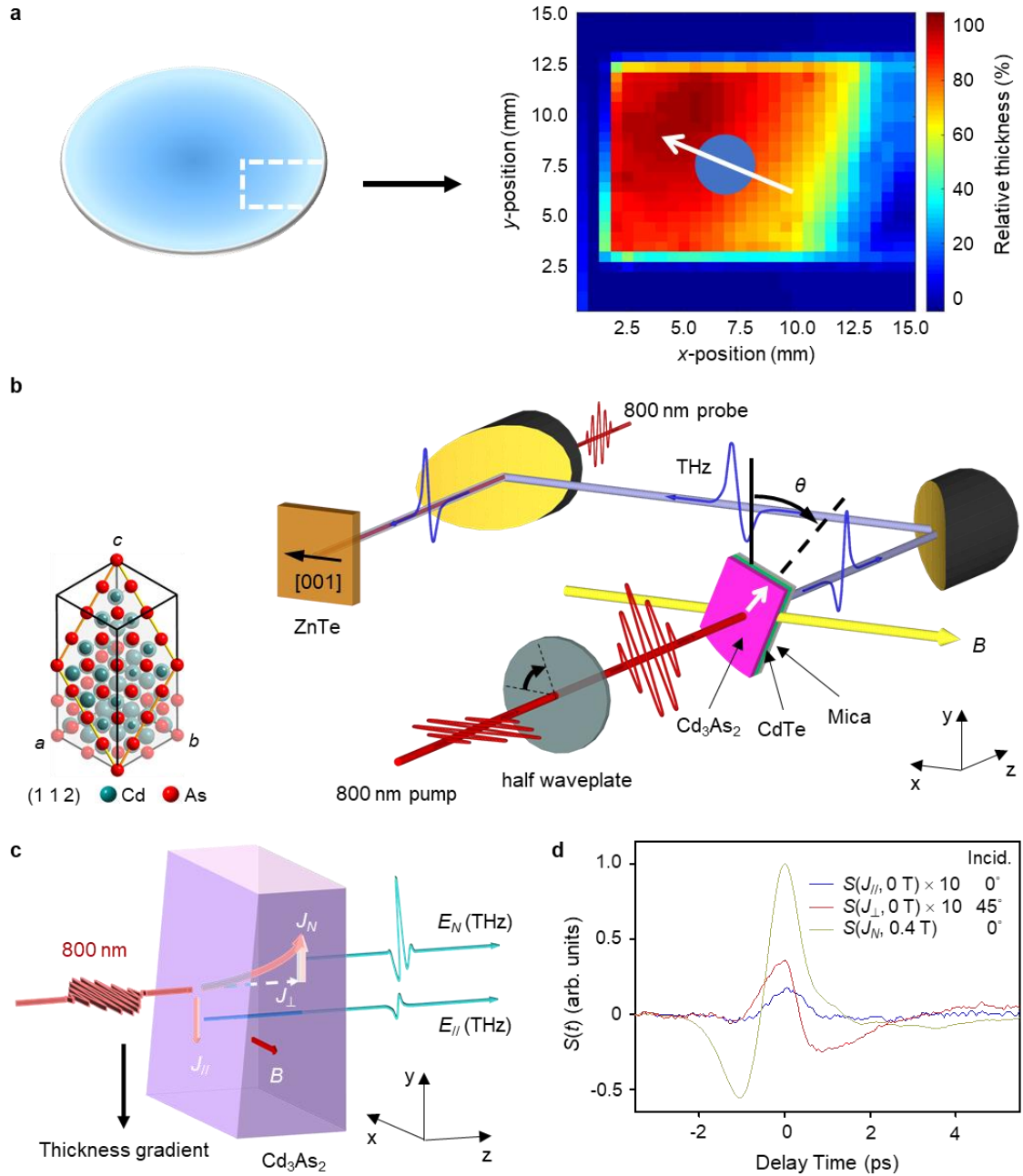

**Fig. 1 THz emissions from Cd$_3$As$_2$ epitaxial film with a thickness gradient. a** The position of a cut piece (white dash line region) in an as-grown wafer, and the mapping of relative thickness of the Cd$_3$As$_2$ film determined by two-dimensional scanning of laser transmittance. The blue circle marks the excitation laser spot and the arrow illustrates the direction of thickness gradient on the spot. **b** Schematic diagram of experimental setup, where $\theta$ denotes the azimuthal angle of the sample with respect to $y$-axis. Inset: the crystal structure of Cd$_3$As$_2$ along (112) plane. **c** Schematic diagram of THz emissions from transient current generated by 800-nm optical excitation. $E_N$ and $E_{//}$ here denote the instantaneous electrical field of THz emission from Nernst current ($J_N$) under magnetic field and from the in-plane current ($J_{//}$), respectively. $J_\perp$ denotes the out-of-plane current. **d** Typical THz waveform generated by $J_{//}$

and $J_\perp$ in (**c**). The two THz waveforms from $J_\perp$ are obtained by sample tilting and by applying an in-plane magnetic field respectively.

**THz emissions from a $Cd_3As_2$ film with a thickness gradient.** The $Cd_3As_2$ samples measured in this work are [112]-oriented thin films grown by molecular beam epitaxy[28-30]. As shown in our previous study[28], the Fermi level of $Cd_3As_2$ is about 250 meV above the Dirac node, the electron doping density is on the order of $10^{13}$ cm$^{-2}$, and the mobility is about $5 \times 10^3$ cm$^2$ V$^{-1}$ s$^{-1}$ at room temperature. The $Cd_3As_2$ film has a radial thickness gradient due to the nature of growth method. The central area is thicker, and the thickness gradually decreases from the center to the outer part on a circular substrate. The gradient is characterized by relative transmittance scanning with a 633-nm He-Ne laser. Figure 1a shows the thickness variation of a $Cd_3As_2$ sample deducted from the transmittance measurement. The maximum thickness of the $Cd_3As_2$ film is estimated to be around 50 nm. The relative thickness gradient is about 10% per millimeter.

Because the $Cd_3As_2$ has strong light absorption at the 800-nm excitation wavelength (~50% for 50-nm thick film[31]), the light intensity decays quickly when penetrating into $Cd_3As_2$, which builds up transient temperature gradient of absorbed light energy and thus the electron temperature gradient along the direction perpendicular to the surface. On the other hand, as a result of the radial thickness gradient of $Cd_3As_2$ film, the ultrafast pulse excitation can also produce a temperature gradient along the in-plane direction. This effect has been previously experimentally verified in $Bi_2Te_3$-based TE thin films[16]. The temperature gradient, along both out-of-plane and in-plane directions, can generated an out-of-plane ($J_\perp$) and in-plane ($J_{//}$) TE current through Seebeck effect, respectively. In our previous studies[11, 28], we have already demonstrated a PTE current in picosecond timescale generated in $Cd_3As_2$ when excited by ultrafast pulses. The picosecond currents emit electromagnetic waves in THz frequency, which can be detected in the electro-optical (E-O) sampling geometry.

The measurement of THz emission from ultrafast laser excitation of $Cd_3As_2$ is schematically depicted in Fig. 1b, which is based on a typical E-O sampling using [110] orientated ZnTe crystal. In a default experimental configuration at room temperature, an 800-nm 150-femtosecond laser beam excites $Cd_3As_2$ film under normal incidence with a magnetic field (when applicable) applied along the *x*-axis, and the E-O sampling is configured to detect *y*-polarized THz signal (with ZnTe[110] along *x*-axis). More experimental details of transient THz emission setup can be found in the method section and the detail of THz sampling geometry can be found in Supplementary Note 2.

As shown in Fig. 1c, the THz field emitted by the in-plane TE current ($J_{//}$) is determined by the following relation: $\mathbf{E}_{//}(t) \propto d\mathbf{J}_{//}/dt$, which propagates along the light path (*z*-axis). The THz field emitted by the out-of-plane TE current $J_\perp$ cannot be detected in the E-O sampling geometry. However, when the sample is tilted, $J_\perp$ would have projection perpendicular to *z*-axis, which contributes to detectable THz emission (see Supplementary Note 3). Alternatively, when an *x*-direction magnetic field is applied, a *y*-direction current $J_N$ is generated from Nernst effect as determined later on. Both approaches can emit detectable THz in the aforementioned E-O sampling geometry.

Figure 1d shows typical THz emission signals detected in E-O sampling with and without magnetic field. According to the magnitudes of THz emission signals ($S(J_{//})$ and $S(J_\perp)$) emitted from $J_{//}$ and $J_\perp$ respectively, we can find that $J_\perp$ is stronger than $J_{//}$ when no magnetic field is applied. Furthermore, when an x-direction magnetic field of $B = 0.4$ T is applied, the THz emission signal ($S(J_N, 0.4\ T)$) for normal incidence is an order of magnitude larger than that for 45°-incidence excitation at 0 T ($S(J_\perp, 0\ T)$), indicating a Nernst effect that is much larger than the Seebeck effect as result of ambipolar transport and possibly anomalous Nernst effect, which will be discussed in the discussion section.

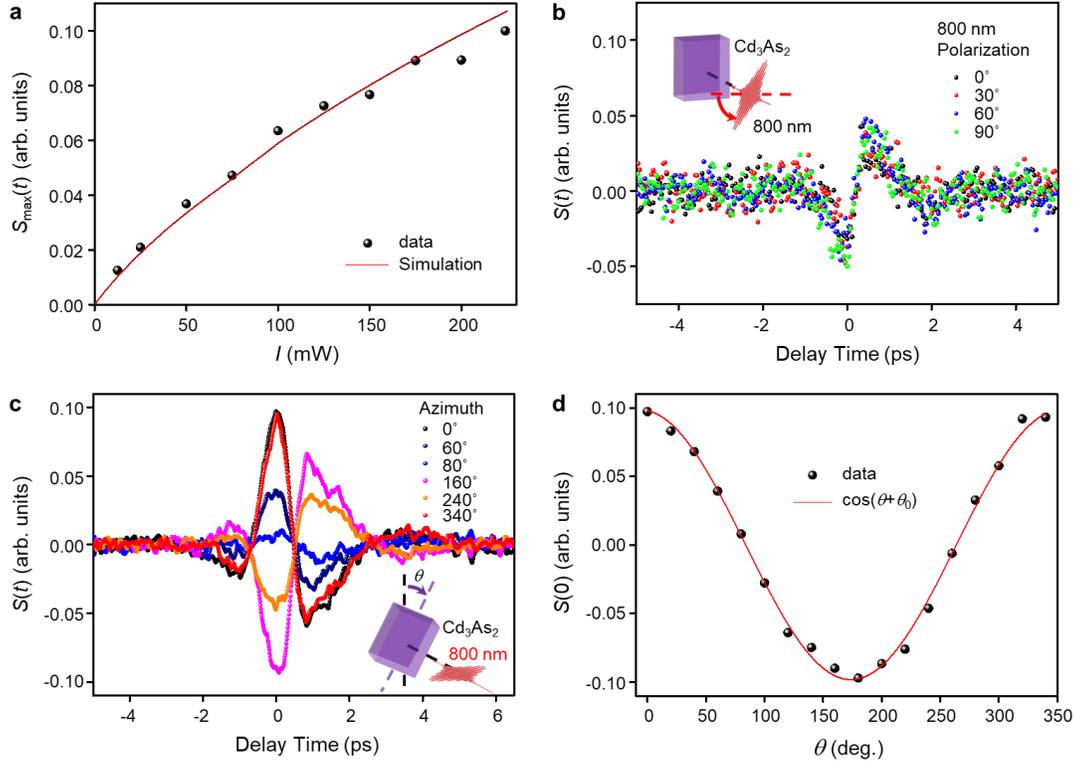

**Fig. 2 THz emission from Cd$_3$As$_2$ without applying magnetic field. a** The laser intensity dependence of the peak field of emitted THz pulse. The red line is the simulation of power dependence according to the TE response presented in Supplementary Note 5. **b** Typical THz waveforms excited by laser with different polarized direction. **c** Typical THz waveforms at different azimuthal angle $\theta$, the thickness gradient is roughly set along y-axis for $\theta = 0°$, with unintentional deviation of $\theta_0$. **d** The azimuthal angle dependence of the peak field of emitted THz pulse, and the red line is the cosinoidal fitting cos ($\theta+\theta_0$), the deviation ($\theta_0$) is 6.2° according to the fitting. Insets in panels (**b**, **c**): the schematic illustrations of the corresponding measurements.

**THz emission from photo-Seebeck effect.** First, we study the characteristics of THz emission from Cd$_3$As$_2$ without applying magnetic field. Figure 2a shows the peak amplitude of THz waveform ($S_{max}(t)$, the waveforms are shown in Supplementary Fig. 4) increases with the excitation intensity ($I$). It can be fitted by a model of PTE effect induced current (see Supplementary Note 5), which approximately follows a power law: $E_{THz} \propto I^{0.66}$. This intensity dependence is different from that of an optical rectification effect or any other second-order effects[17, 32-41], which typically exhibits linear excitation intensity dependence. Here we can rule out the effect from absorption saturation of the excitation beam, because a power-dependent transmittance measurement of the 800-nm excitation beam is flat within the power range as shown in Supplementary Fig. 4a. Furthermore, the THz emission is independent of the polarization of excitation beam as shown in Fig. 2b. This is consistent with the isotropic optical properties on $Cd_3As_2$ (112) plane for near-infrared photon[42-44]. However, if the excitation polarization is fixed and the sample is rotated around the surface normal axis (varying the azimuthal angle $\theta$, as shown in the inset of Fig. 2c), the amplitude of the THz waveform changes with $\theta$ as shown in Fig. 2c. The peak signal of THz emission exhibits a cosinoidal dependence on $\theta$ as shown in Fig. 2d.-According to the cosinoidal fitting shown in Fig. 2d, we find the THz polarization direction is along the direction of thickness gradient of $Cd_3As_2$ film with accuracy of 6.2°. This azimuthal angle dependence of THz emission is consistent with that the THz emission is generated via the in-plane Seebeck current $J_{//}$ induced by thickness gradient[16] as discussed above.

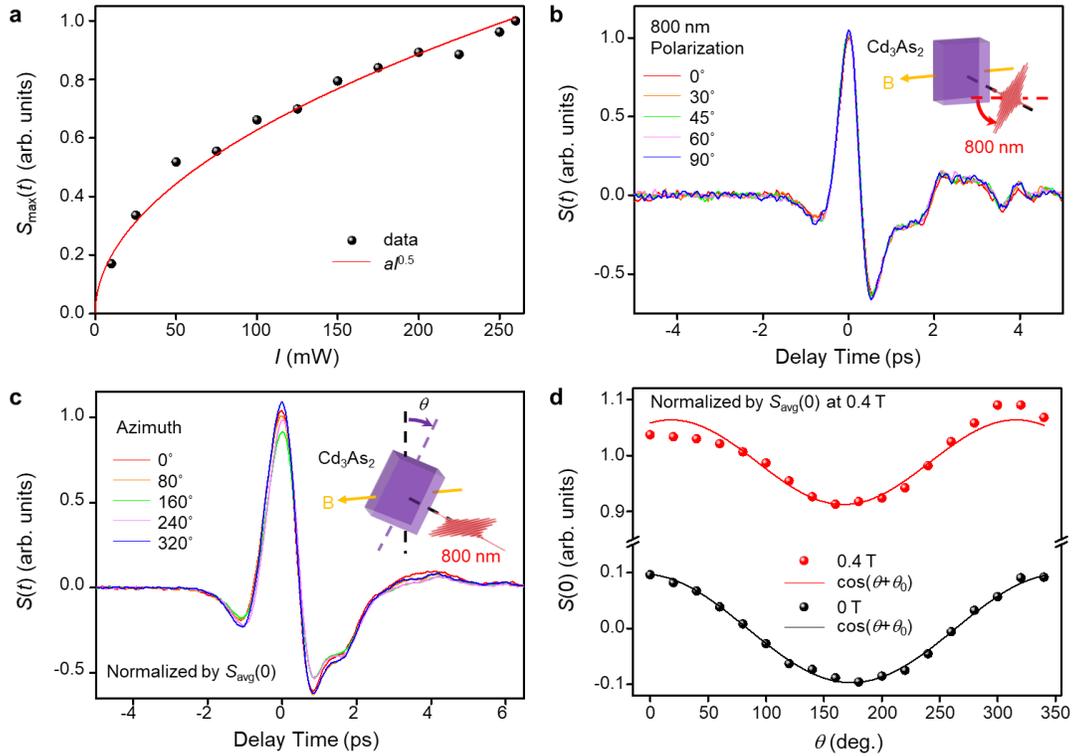

**Fig. 3 THz emission from $Cd_3As_2$ with a magnetic field of ~0.4 T at *x*-direction. a** The laser intensity dependence of the peak field of emitted THz pulse. The red line is a power law fitting. **b** Typical THz waveforms excited by laser with different polarization. **c** Typical THz waveforms at different azimuthal angle $\theta$. **d** The azimuthal angle dependences of the peak field of THz pulses with and without magnetic field, and the lines are the cosinoidal fitting. The data are normalized by the average amplitude at time zero ($S_{avg}(0)$, 0.4 T). Insets in panels (**b**, **c**): the schematic illustrations of the corresponding measurements.

**THz emission by photo-Nernst effect.** In the next, we study the characteristics of the THz emission measurement results when a 0.4-T in-plane magnetic field is applied. After applying the magnetic field along $x$-direction, the THz emission is about an order of magnitude larger. Figure 3a shows the excitation intensity dependence of THz emission amplitude in the presence of a magnetic field. Different from that at 0 T which follows power law of $E_{THz} \propto I^{0.66}$, the excitation intensity dependence under magnetic field follows a different power law of $E_{THz} \propto I^{0.5}$. The THz emission is also independent of the polarization of excitation beam as shown in Fig. 3b. Both power and polarization dependence support the response is due to photothermal effect instead of nonlinear optical effect, which is similar to the case without magnetic field. Figure 3c and 3d show the dependence of the THz waveforms at different azimuthal angle $\theta$. The THz response at 0.4 T magnetic field is composed of a $\theta$-dependent cosinoidal oscillation component and a $\theta$-independent component. The $\theta$-independent component is one order of magnitude larger than the $\theta$-dependent component. The $\theta$-dependent component is from $J_{//}$, the oscillation amplitude of the $\theta$-dependent component is about the same for those taken with and without magnetic field as shown in Fig. 3d. The $\theta$-independent component is induced by the magnetic field due to photo-Nernst effect.

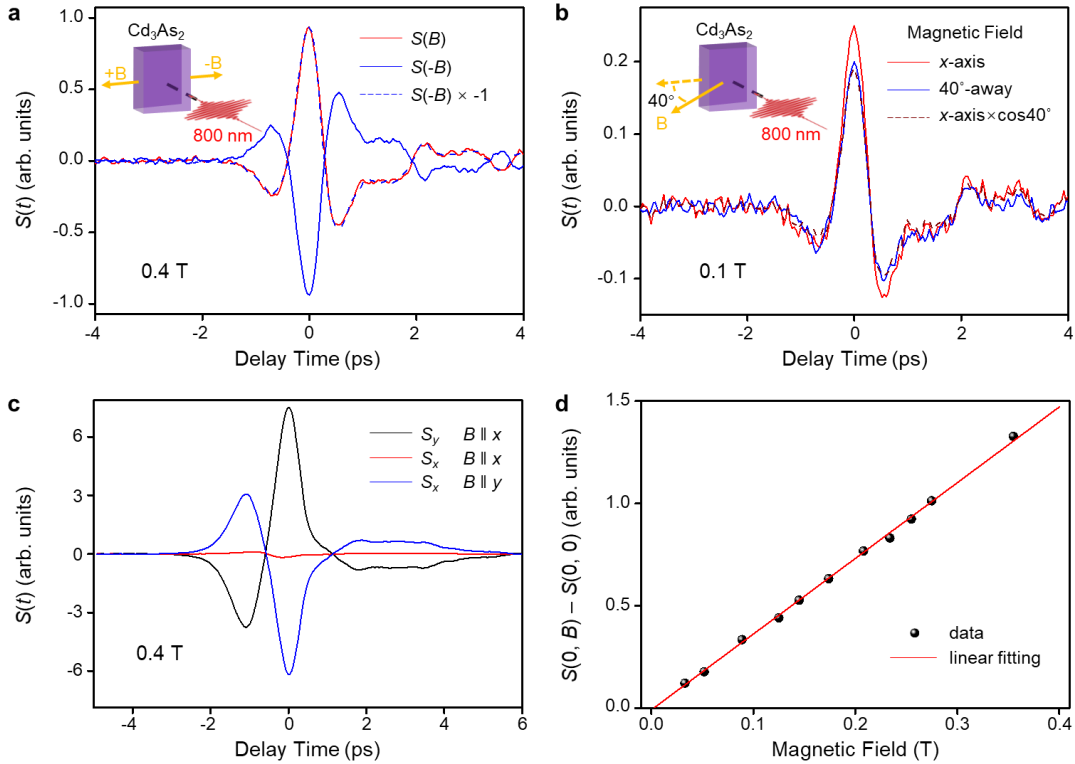

**Fig. 4 Magnetic field dependences of THz emission. a** The THz emissions at reverse magnetic fields of 0.4 T. The dash line is the reversion of waveform at $B = -0.4$ T. **b** The THz emissions when a 0.1 T-magnetic field is applied at $x$-axis and 40° away from $x$-axis in $xz$-plane, respectively. The dash line is the THz waveform under $x$-directional magnetic field multiplying by cos40°. **c** The $x$- and $y$-polarization components of the THz emissions with magnetic field $B\|x$ and $B\|y$. **d** The amplitude of net magnetic THz signal at time zero as a function of the intensity of magnetic field. Insets in panels (**a**, **b**): the schematic illustrations of the corresponding measurements.

In the following, we systematically studied the magnetic field dependence of THz emission. First, we minimized the amplitude of nonmagnetic signal by adjusting $\theta$, therefore only the magnetic field dependent THz component can be detected. We find the THz waveform can flip over and remain the same amplitude as the direction of the magnetic field is reversed as shown in Fig. 4a. However, if $\theta$ is set so that there is a strong nonmagnetic signal, the directly detected THz signals are obviously asymmetric under reverse magnetic fields, unless the nonmagnetic component of the signal is subtracted (Supplementary Fig. 5). This further confirms that the THz emission under magnetic field contains a pure magnetic component ($S(t, B) - S(t, 0)$) and a nonmagnetic component ($S(t, 0)$) that are independent of each other, and the direction of magnetic component is affected by the direction of magnetic field. Moreover, when the magnetic field is 40°-away from the $x$-axis in $xz$-plane, the THz amplitude approximately equals to the THz amplitude under in-plane magnetic field multiplying by cos40°, as show in Fig. 4b. It implies that only the in-plane projection of the magnetic field influences the detected THz emission.

The polarization of THz emission can be determined by measuring the orthogonal polarization THz components $S_x$ and $S_y$ along $x$- and $y$-axis, respectively (see detailed method in Supplementary Note 3). As shown in Fig. 4c, for $B\|x$, $S_y$ dominates the signal and $S_x$ is close to zero, so the THz emission is $y$-polarized; for $B\|y$, the magnetic THz becomes $x$-polarized. We can summarize that the polarization of THz emission is perpendicular to the in-plane magnetic field. The residual signal of $S_x$ at $B\|x$ and the THz amplitudes difference between $B\|x$ and $B\|y$ should be due to experimental errors, eg. ZnTe [001] or $B$ not exactly along $x/y$-axis, and the inhomogeneity of ZnTe crystal. Practically, these errors are difficult to be removed completely. Finally, as shown in Fig. 4d, the pure magnetic THz component is linearly proportional to the in-plane magnetic field (THz waveforms presented in Supplementary Fig. 6).

**Discussion**

The experimental evidences described above are consistent with the interpretations that the observed THz emission is dominated by transient PTE current. Due to the thickness gradient of the sample and strong absorption to the excitation light, the electron temperature gradient is generated after photoexcitation along both in-plane and out-of-plane directions which are responsible for detectable THz emission without and with magnetic field respectively. According to the above experimental results, the THz emission polarized along $y$-axis, which is detectable in the E-O sampling configuration, can be given by the following empirical equation:

$$\mathbf{E}_{\text{THz}} \propto I^{0.66}\hat{\mathbf{y}} + bI^{0.5}(\hat{\mathbf{z}} \times \mathbf{B}) \quad (1)$$

where $b$ is a constant that corresponds to the ratio of Seebeck and Nernst contributions. The first term represents the contribution of THz emission generated from a transverse Seebeck effect due to thickness gradient that is independent of the magnetic field (as shown in Fig. 5a); the second term represents the contribution from the Nernst effect that is dependent on the magnetic field. These two terms exhibit different power law dependences according to Fig. 2a and 3a, and the second term has a linear dependence on $B$ according to Fig. 4d. The direction of the TE current is determined by the thermal gradient of the sample, which does not rely on the polarization of excitation light (as observed in Fig.

2b and 3b).

The sublinear excitation intensity dependence and the independence of the excitation polarization help to rule out other nonlinear optical effects which are usually considered as the mechanisms of THz emission. These effects include optical current injection, shift current, optical rectification, depletion field effect, photo-Dember effect and so on[15, 17, 33-35, 39, 41]. Although the second order nonlinear tensor is zero due to inversion symmetry of bulk $Cd_3As_2$, nonzero third order tensor can exist around an interface or by the assistance of a perpendicular DC electric field. However, those second order nonlinear optical responses should have linear power dependence instead of sublinear power dependence. The deduction of sublinear power dependence of the TE response, either the transvers Seebeck effect or the Nernst effect, is presented in Supplementary Note 5, by assuming that the intensity of transient currents is dominated by the laser induced TE effects. The polarization independence is also consistent with the polarization independent absorption of $Cd_3As_2$.

In the next, we analyze the TE response more quantitatively. The transient thermal electric current along y-axis ($J_y(B)$), which emits THz emission that is detectable when setting [001] of ZnTe crystal along x-axis, is composed of two components:

$$J_y(B) = J_{S//}(B) + J_N(B) = \alpha_{yy}(B)\frac{dT_e}{dy} + \alpha_{yz}(B)\frac{dT_e}{dz} \tag{2}$$

where $J_{S//}$ and $J_N$ are the transverse Seebeck and the Nernst currents respectively, $T_e$ is the instantaneous electron temperature, and $\alpha_{yy}$, $\alpha_{yz}$ are the TE conductivities of Seebeck and Nernst effect. When a magnetic field $B$ is applied along $x$ direction, the TE conductivities can be written as: $\alpha_{yy} = \alpha_0 \frac{1}{1+(\mu B)^2}$ and $\alpha_{yz} = \alpha_0 \frac{\mu B}{1+(\mu B)^2}$, as derived from the linearized Boltzmann equation[45], where $\alpha_0$ is the zero field TE conductivity and $\mu$ is the carrier mobility. Taking $\mu$ ~5000 cm$^2$ V$^{-1}$ S$^{-1}$ according to our previous study[28], and $B$ ~0.4 T in the measurement, $(\mu B)^2$ ~ 0.04 << 1. Therefore,

$$J_y(B) \approx \alpha_0 \frac{dT_e}{dy} + \mu B \alpha_0 \frac{dT_e}{dz} \tag{3}$$

which is fully consistent with empirical Eq. 1 summarized from the experimental observation. According to Eq. 3, the second term (Nernst current $J_N$) should be smaller than the longitudinal Seebeck current $J_{S\perp} = \alpha_0 \frac{dT_e}{dz}$, this is inconsistent with the experimental observation that $J_N$ is over an order of magnitude larger than $J_{S\perp}$ as shown in Fig. 1d and Supplementary Fig. 2d.

This large Nernst signal is due to the ambipolar transport nature during the transient process as illustrated in Fig. 5b. To describe the Nernst response, Eq. 3 would become $J_y(B) \approx \alpha_{e0}\frac{dT_e}{dy} - \alpha_{h0}\frac{dT_h}{dy} + \mu_e B \alpha_{e0} \frac{dT_e}{dz} + \mu_h B \alpha_{h0} \frac{dT_h}{dz}$, where $e$ and $h$ denote the electron and hole respectively. In the right side of this equation, the first two terms are contributions from electron and hole to $J_{S//}$ respectively, and the last two terms are contributions to $J_N$. These two types of carriers have the same longitudinal thermal diffusion direction, counterbalancing each other's contribution to the Seebeck current. However, distinct from the Seebeck effect, the electrons and holes are deflected in opposite directions under magnetic field, and thus, their contributions to the Nernst current are added together as shown in Fig. 5b. The lifetime scale of photoexcited electrons and holes are consistent with recent experimental observation that the photoexcited electrons and holes have an approximate lifetime of 3 ps before recombination in a Tr-ARPES measurement[46]. The picosecond photoexcited carrier lifetime is sufficient to support the transient ambipolar transport on picosecond timescale and the subsequently enhanced THz emission. Usually, it is not easy to observe the ambipolar transport in low temperature electrical transport measurement, because one type of carriers predominated in heavily doped semimetals. Previously, the Nernst signal enhancement had been observed in compensated semimetals such as bismuth and graphite hosting almost equal concentrations of electrons and holes under transport measurement without photoexcitation[45, 47, 48]. The transient PTE response and suitable photoexcited carrier lifetime of $Cd_3As_2$ provide an ideal experimental platform to observe the enhancement on Nernst effect as demonstrated in this work.

Additionally, the anomalous Nernst effect may also contribute to the large enhancement of the PTE signal under magnetic field[49, 50]. The anomalous Nernst effect arises from the nonzero Berry curvature ($\Omega_B$) associated with the Weyl nodes without the need to apply a real magnetic field. However, it should be absent in Dirac semimetal without applying magnetic field due to restoration of time-reversal symmetry. When a magnetic field in applied, the magnetic field can split a Dirac node into a pair of Weyl nodes with opposite chirality, and then a nonzero Berry curvature is generated. It has been found that the Berry curvature can induce an anomalous Nernst effect to add on the conventional one[6, 49-51]. At low magnetic field and low temperature, it is predicted theoretically that the total Nernst effect is characterized by an almost steplike profile at low magnetic field and the response is dominated by the anomalous Nernst effect[49, 50]. Although in this work, we cannot clearly sort out the contribution from conventional and anomalous Nernst effects, the large enhancement factor implies very possible contribution from anomalous Nernst response, which calls for further studies.

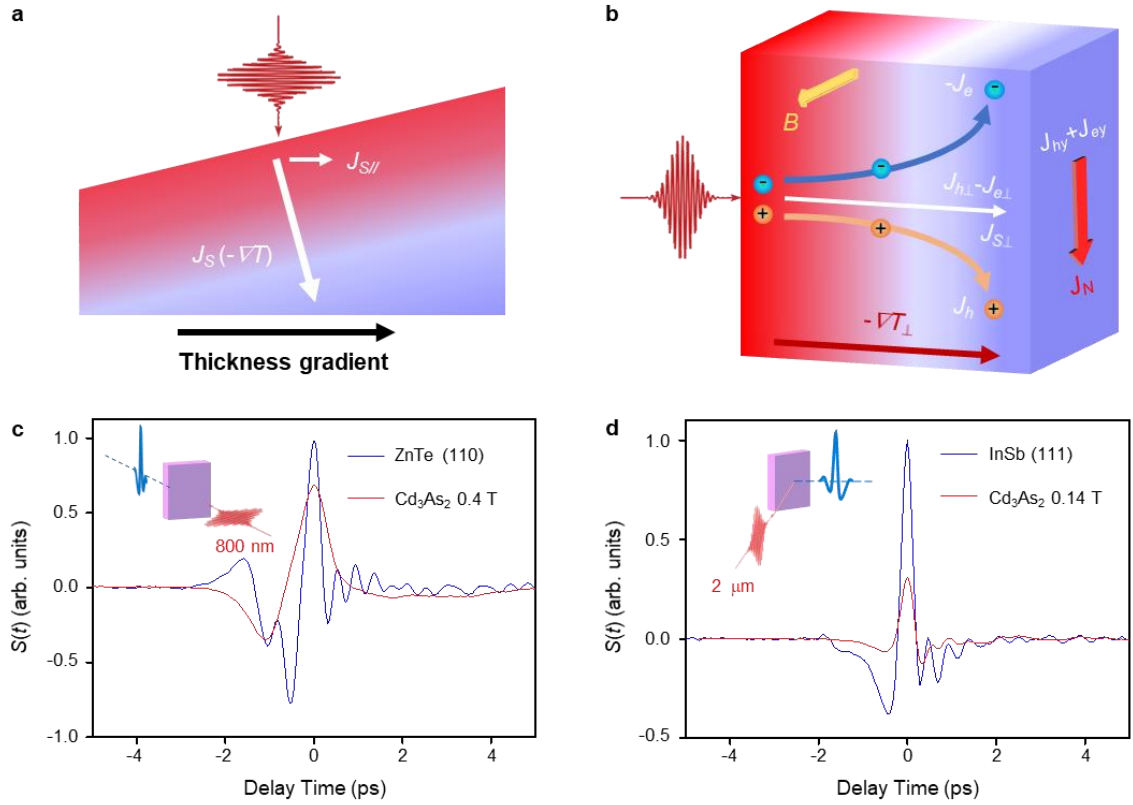

**Fig. 5 Mechanisms of PTE THz emissions and the comparison with semiconductor THz sources.** Schematics of (**a**) transverse Seebeck effect and (**b**) Nernst effect, where $J_S$ is the Seebeck current, $J_{S/\!/}$ is the in-plane component, $J_{S\perp}$ and $J_N$ denote the out-of-plane component and the Nernst current respectively, which are composed by both the electron current and hole current. Comparison of the optimized THz emission from $Cd_3As_2$ nano-films with the (**c**) ZnTe (110) and (**d**) InSb (111) wafers with thicknesses of 0.5 mm, respectively. Insets in panels (**c, d**): the corresponding setups.

At last, we compared the THz emission from $Cd_3As_2$ by photo-Nernst effect at weak magnetic field with that from the typical semiconductor THz sources. Figure 5c compares the THz intensity from the response with ZnTe, which is the most widely used THz crystal through optical rectification[52]. The intensity of THz emission from $Cd_3As_2$ at 0.4 T can reach ~70% of that from 0.5 mm-thick ZnTe optimized under the same excitation condition. Figure 5d further compare the THz emission from $Cd_3As_2$ with that from a 0.5 mm-thick InSb (111), which is one of the best narrow bandgap semiconductors for THz source with the dominated mechanism of photo-Dember effect[34, 35, 53]. The intensity of THz emission from $Cd_3As_2$ at 0.14 T can reach ~30% of that from InSb under a reflective detection geometry with 2-μm excitation, confirming the thermoelectric THz emission can also be triggered by longer wavelength due to the zero-bandgap nature of $Cd_3As_2$. The comparison measurements indicate that, based on the photoexcited transient Nernst effect, the THz emission efficiency from $Cd_3As_2$ with thickness of tens nanometer is already comparable with typical semiconductor THz sources with submillimeter thickness. When integrated with ferromagnetic materials with careful design[54-58], it would provide a highly efficient compact THz source solution based on photo-Nernst effect.

In summary, we have studied THz emissions generated from transient PTE current in $Cd_3As_2$ by femtosecond laser excitation under zero and weak magnetic field at room temperature. The THz emissions are generated from a thickness-gradient-assisted transverse Seebeck and Nernst effect, respectively. The magnitude of photo-Nernst current is an order of magnitude larger than Seebeck current at low magnetic field indicating ambipolar transport nature after the photoexcitation. Because of the Nernst effect, THz emission efficiency is comparable to ZnTe and InSb crystals. The THz emission of $Cd_3As_2$ can be triggered by ultrafast laser with wavelength over broad spectrum range due to its semimetallic nature. In addition, the THz emission can be versatility controlled through optical, electric, magnetic and thermal approaches. Compared with its two-dimensional Dirac semimetal counterpart, graphene, which has shown strong and fast PTE responses[10, 25, 59], $Cd_3As_2$ has larger steady state TE coefficients and smaller thermal conductivity[49, 60-64], which are both more favorable for TE conversion than graphene and thus promises better PTE response at transient state. We expect the transient PTE dynamics revealed in this work provides indispensable device physics for high speed/field electronic and optoelectronic device applications based on three dimensional topological semimetals.

**Methods**

**Sample preparation.** $Cd_3As_2$ thin films with about 50-nm thickness were grown in a PerkinElmer (Waltham, MA) 425B molecular beam epitaxy system. Fresh cleaved 2-inch mica was used as the substrate. The data presented in the maintext are obtained from two samples with substrate thickness of 30 µm and 70 µm respectively. The absorbed molecules on the surface of substrate were removed by annealing at 300 °C for 30 min. Then a CdTe buffer layer with a thickness of approximately 10 nm was deposited on the substrate to assist $Cd_3As_2$ nucleation. After that, the $Cd_3As_2$ layer was grown on the buffer layer by evaporating the bulk material (99.9999%, American Elements Inc., Los Angeles, CA) at 170 °C. The growth was *in situ* monitored by the reflection high-energy electron diffraction system. The $Cd_3As_2$ layer has high quality and excellent electric properties, which have been systematically studied in previous works[28-30]. The relative thickness mapping of $Cd_3As_2$ film is determined by two-dimensional transmittance scanning with a 633-nm cw laser. The laser spot size is about 1 mm, and the scanning step is 0.5 mm.

**Transient THz emission setup.** In a default experimental configuration, 100-fs laser pulses with a central wavelength of 800 nm from a 250-kHz Ti-sapphire amplifier (RegA) system[65] were used to excite the samples and to probe the emitted THz pluses through a 0.5 mm-thick ZnTe (110) crystal using a standard E-O sampling technique[66]. The excitation beam propagates along the *z*-direction. It passes through a 3-mm-diameter aperture to shape a circular excitation beam, and the full width at half maximum is about 1.6 mm. The excitation power is measured just in front of the sample and chopped with a frequency about 1 kHz. A half waveplate is mounted into the setup to adjust the polarization direction of the excitation beam before incident onto a sample. The incident angle could be adjusted by tilting the sample in *xz*-plane, and it is defined as 0° for normal incidence on the $Cd_3As_2$ side. The THz pulse was emitted along the *z*-direction, and then it was collected and focused onto the ZnTe crystal using a pair of parabolic mirrors, which is collimated with a *y*-polarized sampling beam. Because the E-O sampling is THz polarization dependent[67], in general measurements, the ZnTe [001] axis is set along *x*-direction to detect *y*-polarized THz component. To acquire the *x*-component, ZnTe [001] axis could be set along *y*-directions. In magneto measurements, a pair of permanent magnets was applied along *x*-direction. The intensity of the magnetic field could be controlled by change the spacing between the two magnets. The maximum magnetic field can reach 0.4 T in this setup. In out-of-plane magnetic field measurement, the magnetic field is applied with an approximately 40°-angle with respect to the *x*-axis in *xz*-plane, and the intensity is lowered to 0.1 T to get a larger space between the permanent magnets to permit the unhindered passing through of the THz wave. In comparative measurement with InAs (111), a reflective setup was used and the excitation beam with a wavelength of 2 μm was from an optical parametric amplification. The light is incidence on the 45°-tilted sample along *x*-direction. the magnetic field can only reach 0.14 T and the direction of the magnetic field has to set along *y*-direction as limited by the reflection geometry. All measurements were carried out in atmosphere condition at room temperature.

**Data availability**

The source data underlying Figs. 1a, d, 2, 3, 4, and 5c, d are provided as a Source Data file [https://doi.org/10.6084/m9.figshare.18482018]. Other data is available from the corresponding author upon reasonable request.

**Acknowledgments**
This project has been supported by the National Key Research and Development Program of China (Grant No. 2020YFA0308800, 2021YFA1400100), the National Natural Science Foundation of China (Grant No. 12034001). D.S. acknowledges Beijing Nature Science Foundation (Grant No. JQ19001). F.X. acknowledges the supports from the Science and Technology Commission of Shanghai (Grant No. 19511120500), the Shanghai Municipal Science and Technology Major Project (Grant No. 2019SHZDZX01) and the Program of Shanghai Academic/Technology Research Leader (Grant No. 20XD1400200). J.C. acknowledges support from the Scientific Research Project of the Chinese Academy of Sciences (Grant No. QYZDB-SSW-SYS038). J.Lai is also supported by China Postdoctoral Science Foundation (Grant No. 2021M690231) and China National Postdoctoral Program for Innovative Talent (Grant No. BX20200015).

**Author contributions:**
D.S. and W.L. conceived the project. L.W., Z.F., J.M., J.Lai, X.S., X.Z. and Z.X. performed the optical experiments under the supervision of D.S. Y.Y. prepared the $Cd_3As_2$ samples and performed the basic characterizations under the supervision of F.X. J.C., S.Z., J.Liu and X.H. contributed to the results analyzation. W.L. and D.S. wrote the paper, assisted by Z.F., J.M., X.Z. and Z.X. All the authors commented on the paper.

**Competing interests**
The authors declare no competing interests.

**Additional information**
Supplementary information is available for this paper at https://doi.org/xxxxx

# Supplementary Information
# Ultrafast photothermoelectric effect in Dirac semimetallic Cd$_3$As$_2$ revealed by terahertz emission

Wei Lu[1,2], Zipu Fan[2], Yunkun Yang[3], Junchao Ma[2], Jiawei Lai[2], Xiaoming Song[1], Xiao Zhuo[2], Zhaoran Xu[2], Jing Liu[1], Xiaodong Hu[1], Shuyun Zhou[4,5], Faxian Xiu[3], Jinluo Cheng[6] & Dong Sun[2,5,*]

[1]State Key Laboratory of Precision Measurement Technology and Instruments, School of Precision Instruments and Opto-electronics Engineering, Tianjin University, Tianjin 300072, China

[2]International Center for Quantum Materials, School of Physics, Peking University, Beijing 100871, China

[3]State Key Laboratory of Surface Physics and Department of Physics, Fudan University, Shanghai 200433, China

[4]State Key Laboratory of Low Dimensional Quantum Physics and Department of Physics, Tsinghua University, Beijing, 100084, China

[5]Collaborative Innovation Center of Quantum Matter, Beijing 100871, China

[6]Changchun Institute of Optics, Fine Mechanics and Physics, Chinese Academy of Sciences, Changchun 130033, China

*Corresponding author. Email: sundong@pku.edu.cn (D.S.)

**Supplementary Note 1. Thermoelectric coefficients of different materials**

Supplementary Table 1 summarize the Seebeck ($S_{xx}$) and Nernst ($S_{xy}$) coefficients of typical topological semimetals, conventional metallic and semiconducting thermoelectric materials, and materials with Nernst coefficients that are comparable with Seebeck coefficients, which are usually called "giant Nernst effect materials". We note many topological semimetals are excellent thermoelectric materials and thermoelectricity serves as one of their most important potential applications.

**Table 1. Seebeck ($S_{xx}$) and Nernst ($S_{xy}$) coefficients of typical topological semimetals, conventional thermoelectric materials, and giant Nernst effect materials.**

| Materials | $S_{xx}$ (μV/K) | $S_{xy}$ (μV/K) |
|---|---|---|
| Cd$_3$As$_2$ [a] | ~200 (150 K, 0 T)[1] | ~160 (200 K, 10 T)[2] |
| ZrTe$_5$ [a] | ~500 (100 K, 13T)[3] | ~5000 (100 K, 13 T)[3] |
| NbAs [a] | ~450 (100 K, 14 T)[4] | ~500 (140 K, 14 T)[4] |
| NbP [a] | ~1100 (30 K, 12 T)[4] | ~800 (109 K, 9 T)[5] |
| TaAs [a] | ~750 (35 K, 10 T)[4] | ~1800 (60 K, 14 T)[4] |
| TaP [a] | ~1700 (150 K, 0 T)[4] | ~1700 (150 K, 0 T)[4] |
| WTe$_2$ [a] | ~90 (300 K, 0 T)[6] | ~4000 (3.7 K, 17 T)[7] |
| PbS [b] | ~400 (300 K, 0 T)[8] | NA |
| SnSe [b] | ~550 (300 K, 0 T)[9] | NA |
| Mg$_2$Si [b] | ~150 (300 K, 0 T)[10] | NA |
| Au [b] | ~9 (5 K, 0 T)[11] | ~10$^{-4}$ (µV/K T)[12] |
| YbAl$_3$ [b] | ~80 (300 K, 0 T)[13] | ~0.22 (30 K, 2 T)[13] |
| SrTiO$_3$ [b] | ~180 (6 K, 0 T)[14] | ~9 (0.9 K, 10 T)[14] |
| PbTe [b] | 265~510 (300 K, 0 T)[15] | >25 (100 K, 1.5 T)[15] |
| URu$_2$Si$_2$ [c] | ~22 (15 K, 0 T)[12] | ~30 (5 K, 12 T)[12] |
| BaFe$_2$As$_2$ [c] | ~8 (170 K, 0 T)[16] | ~29 (45 K, 30 T)[16] |
| P$_4$W$_{12}$O$_{44}$ [c] | ~70 (15 K, 0 T)[17] | ~13 (15 K, 9 T)[17] |

[a] topological semimetals
[b] conventional thermoelectric materials
[c] giant Nernst effect materials

**Supplementary Note 2. THz detection by electro-optical sampling with ZnTe crystal**

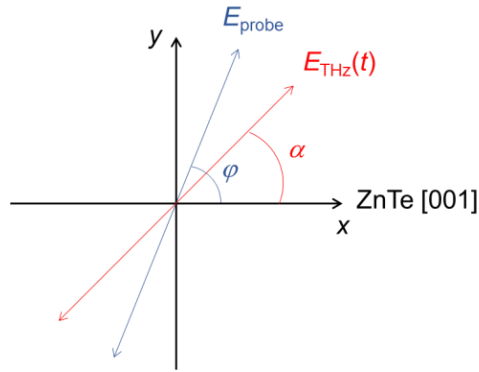

**Supplementary Fig. 1 E-O sampling geometry.** Configuration of the THz electric field ($E_{THz}$) direction and 800-nm sampling beam polarization ($E_{probe}$) direction with respect to the ZnTe [001] axis.

The electro-optical (E-O) sampling signal $S(t, \alpha, \varphi)$ of linearly polarized THz wave detected by standard E-O sampling method using ZnTe (110) crystal can be expressed as[18]:

$$S(t, \alpha, \varphi) \propto E_{THz}(t) \, (2 \sin\alpha \, \cos2\varphi + \cos\alpha \, \sin2\varphi) \quad (1)$$

where $\alpha$ and $\varphi$ are the polarization angles of the THz wave and the 800-nm sampling beam with respect to the [001] direction of ZnTe crystal as shown in Supplementary Fig. 1, $t$ is the time delay of the sampling beam, and $E_{THz}(t)$ is the electric field of THz wave. If the ZnTe [001] axis is along $x$-axis, and the polarization of sampling beam $E_{probe}$ is along $y$-axis ($\varphi = \pi/2$), Supplementary Eq. 1 becomes:

$$S(t, \alpha) \propto 2 \, E_{THz}(t) \sin(\alpha + \pi) = -2 \, E_{THz}(t) \sin\alpha \quad (2)$$

Therefore, the ZnTe crystal is acting like a polarizer and the detected THz signal follows a sinusoidal function of the THz polarization angle. When $E_{THz}$ is along the ZnTe [001] axis, $\alpha = 0$, $S(t, 0) = 0$, which corresponds to no E-O sampling signal. Alternatively, we can align the ZnTe [001] axis along y-axis ($\varphi = 0$) so that x-polarized component of THz wave are maximized in this E-O sampling geometry ($S(t, \alpha) \propto -2 E_{THz} \cos\alpha$).

If the THz wave is elliptically polarized, $\boldsymbol{E}(t) = e^{-i(\omega t+\alpha_0)}(E_x, iE_y)$, where $\alpha_0$ is the initial phase of the elliptical light, which depends on the azimuthal angle ($\theta$) of the sample. Therefore, according to the Supplementary Eq. 2, the peak position of the THz waveform will have a shift when varying $\theta$.

**Supplementary Note 3. Detection of THz emission from in-plane and out-of-plane transient currents**

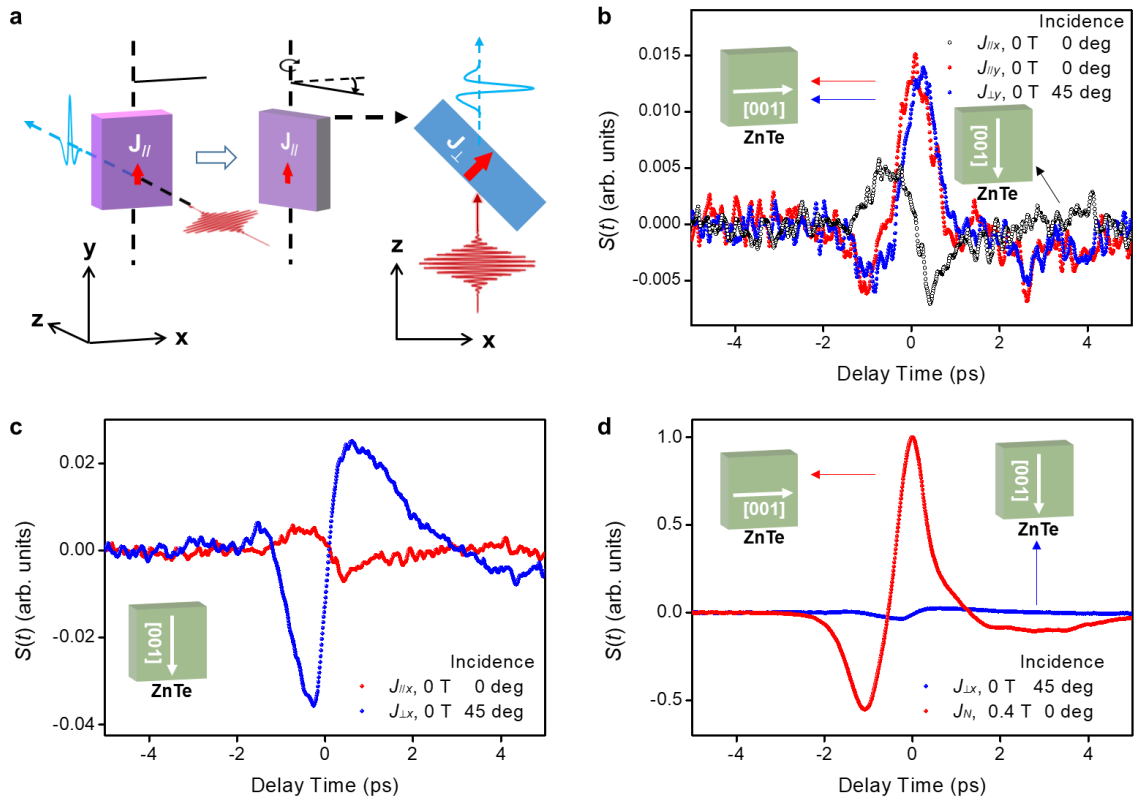

**Supplementary Fig. 2 THz emission from in-plane and out-of-plane transient currents. a** Side view of normal incidence and oblique incidence geometry of excitation beam (left two configurations) and top view of oblique incidence (right configuration). **b** THz emission for normal incidence and 45°-oblique incidence geometry excitation. In the detection geometry, the [001] direction of ZnTe is set on x-axis, so the electric field component of THz emission along y-axis is detected. **c** Then the [001] direction of ZnTe is set on y-axis, so the electric field component of THz emission along x-polarization is detected. **d** THz waveform at 0.4 T for normal incidence with the [001] direction of ZnTe set on x-axis and THz waveform for 45°-incidence excitation at 0 T with the [001] direction of ZnTe set on y-axis. The amplitude of THz signals at 0.4 T is set as 1 in the plot.

As shown in Supplementary Fig. 2a, under normal incidence, only the THz wave emitted by the in-plane transient current ($J_{//}$) can propagate along the light path (along $z$-axis), thus can be detected according to the experimental geometry. The propagation of THz wave emitted by the out-of-plane transient current ($J_\perp$) is perpendicular to $z$-axis under normal incidence, so it cannot be detected in the E-O sampling configuration. When the sample is excited with an oblique angle, $J_\perp$ can have a nonzero projection along $x$-direction ($J_{\perp x}$) and the THz emission from $J_{\perp x}$ can be detected. To distinguish the THz emission contribution from $J_{\perp x}$ under the oblique excitation condition, the $x$-component of the in-plane current $J_{//}$ need to be minimized to avoid overlapping with THz emission signal from $J_{\perp x}$.

**i. In-plane current detection**

First, we set the ZnTe [001] along $x$-axis to detect the $y$-polarized THz wave according to the discussion in Supplementary Note 2. The azimuthal angle of sample is varied to obtain optimum THz signal, so that $J_{//}$ is along $y$-axis to get optimized THz detection efficiency. Then we fixed the ZnTe [001] along $y$-axis to detect the $x$-polarized THz wave. At normal incidence, the $x$-component of $J_{//}$ is significantly smaller than the $y$-polarized component as shown in Supplementary Fig. 2b ($J_{//y}/J_{//x}$ ~2.9), here $J_{//x}$ survives because imperfect optimization of $J_{//y}$ due to the limited THz signal when rotating the azimuthal angle of sample. When the sample is rotated along $y$-axis by 45° (middle configuration of Supplementary Fig. 2a) to have 45°-oblique incident excitation, the THz waveform obtained with ZnTe [001] along $x$-axis almost keeps the same as shown in Supplementary Fig. 2b, because $J_{//y}$ contributes to the detected THz signal in both configurations.

**ii. Out-of-plane current detection by oblique incidence**

The out-of-plane current $J_\perp$ is only detectable when we set the ZnTe [001] along $y$-axis. At 45°-oblique incidence, $J_\perp$ has a component projected onto $x$-axis ($J_{\perp x}$), which provides detectable THz signal as shown in Supplementary Fig. 2c. The peak-to-peak amplitude of $J_{\perp x}$ is about 3 times larger than $J_{//y}$, implying $J_\perp$ is much stronger than $J_{//}$.

**iii. Out-of-plane current detection by applying an in-plane magnetic field**

Alternatively, instead of rotating the sample along $y$-axis to have an oblique incidence, an in-plane magnetic field can be applied to deflect the direction of $J_\perp$ to provide a detectable transient current ($J_N$) in a classical picture, which corresponds to Nernst response in this work. Experimentally, by applying an in-plane magnetic field of 0.4 T, the emitted THz signal is one order of magnitude larger than that for 45°-oblique incidence at 0 T as shown in Supplementary Fig. 2d and maintext Fig. 1d.

**iv. Comparison of THz spectra**

The THz waveforms generated from $J_{//y}$ at normal incidence, $J_{\perp x}$ under 45°-oblique incidence and $J_N$ under normal incidence with 0.4 T magnetic field and their Fourier transform spectra are shown in Supplementary Fig. 3a and 3b respectively. At normal incidence, the THz waveforms for $B$ = 0 T and 0.4 T are similar after normalization as shown in Supplementary Fig. 3a; while at 45°-oblique incidence, the THz waveform is very different, suggesting the transport dynamics may be different for in-plane and out-of-plane transient current. In the frequency domain (Supplementary Fig. 3b), the central frequency of emitted THz spectrum (0.18~0.25 THz) is slightly different for $B$ = 0 T and 0.4 T at normal incidence and the peak of 0.4 T has a flat and broader top. The full widths at half maximum (FWHM) of the three THz pulses are all at about 0.5 THz.

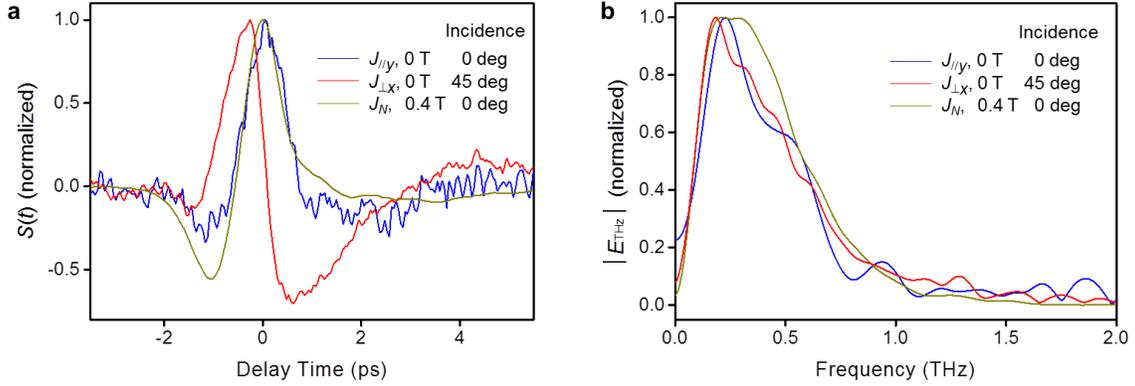

**Supplementary Fig. 3 THz spectra. a** The normalized THz waveforms obtained at a magnetic field intensity of 0 T and 0.4 T respectively. **b** The normalized Fourier spectra.

**Supplementary Note 4. Excitation power dependence of THz emission at $B$ = 0 T**

Under different excitation intensity, the THz waveforms are very similar, but the maximum amplitude has a shift of ~0.3 ps as the laser intensity increases from 12.5 mW to 225 mW as shown in Supplementary Fig. 4b. As a consequence, the frequency spectrum has a blue shift (Supplementary Fig. 4c and 4d), indicating a faster establishment of the transient current when excitation power increases.

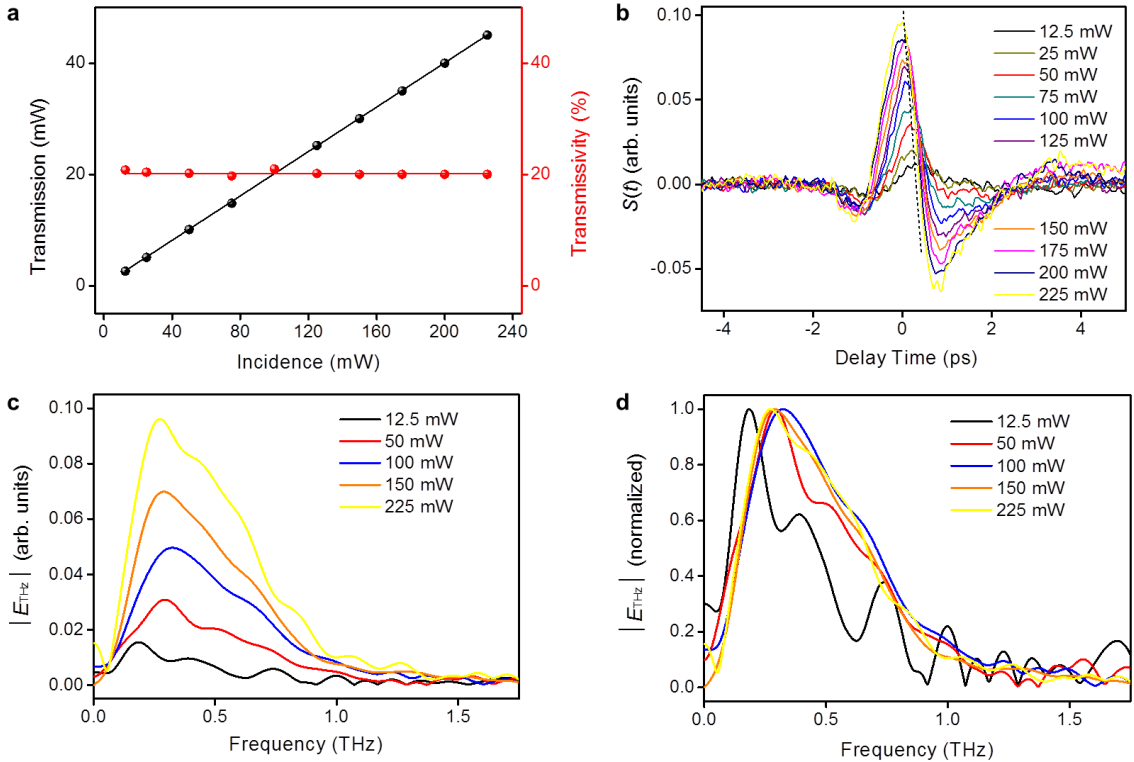

**Supplementary Fig. 4 Excitation power dependent measurements of THz emission at $B$ = 0 T. a** Excitation power dependence of sample transmittance at 800 nm. **b** The THz waveforms measured at different excitation power. **c** Fourier transform spectra for THz waveforms at 5 different excitation power, and (**d**) the normalized spectra of (**c**).

**Supplementary Note 5. Laser intensity dependence of thermoelectric current**

In this session, we try to elaborate the dependence of thermoelectric current on laser intensity for photothermoelectric response. After laser pulse absorption, the photocarriers will be thermalized by rapid carrier-carrier scattering and reach a quasi-equilibrium Fermi-Dirac distribution that can be characterized by instantaneous electron temperature $T_e$. Due to the spatial temperature gradient of $T_e$, there is a thermoelectric current response, which can be written as $J_i = \alpha_{ij} \nabla T_{ej}$. In general, both the coefficient $\alpha_{ij}$ and the temperature gradient $\nabla T_{ej}$ is dependent on the electron temperature and they are discussed separately below.

Firstly, we consider the temperature gradient $\nabla T_e$. Ignoring the thermal transfers to lattice and the environment during the first few picoseconds after the photoexcitation, the increase of instantaneous thermal energy of the electron system induced by laser absorption is

$$\int_{T_0}^{T_e} C_e(T_e) dT_e = \beta(z) I \tag{3}$$

where $T_0$ is the temperature before excitation, which is room temperature in our experiment; $C_e(T_e) = \gamma T_e$ is the electron specific heat, which is proportional to electron temperature; $\beta(z)$ is the relative light absorption at depth of $z$ in Cd$_3$As$_2$ from the illuminated surface and $I$ is the laser fluence. The instantaneous electron temperature can be written as:

$$T_e = \sqrt{\frac{2\beta(z)I}{\gamma} + T_0^2} \tag{4}$$

In our experiment, the magnitude of $\beta(z)$ should be about $10^5$ cm$^{-1}$, $\gamma \sim 70$ μJ cm$^{-3}$ K$^{-2}$ [19], $T_0 \sim 300$ K, the average laser fluence $I$ is in range of 5 ~ 90 μJ cm$^{-2}$ in the laser intensity dependent measurements. $T_e$ can be estimated to be from 40 K (5 μJ cm$^{-2}$) to 400 K (90 μJ cm$^{-2}$) above the room temperature. This estimation is consistent with Tr-ARPES measurement[20]. As the next step, the temperature gradient can be written as:

$$\nabla T_e = \frac{dT_e}{dz} = \frac{I}{\gamma \sqrt{\frac{2\beta(z)I}{\gamma} + T_0^2}} \frac{d\beta(z)}{dz} \tag{5}$$

For the in-plane Seebeck effect described in the maintext, the temperature gradient should follow the similar relation with $I$.

Then we consider the thermoelectric conductivity $\alpha_{ij}$. Without magnetic field, $\alpha_{ij}$ can be written as $\alpha_{ii} = \sigma_{ii} S_{ii}$, where $S_{ii}$ is the Seeback coefficient and $\sigma_{ii}$ is the electrical conductivity. At weak magnetic field limit, both coefficients ($S_{ii}$ and $\sigma_{ii}$) are nearly isotropic[21-23], and the value of $S_{ii}$ and $\sigma_{ii} = \rho_{ii}^{-1}$ can be obtained from transport measurements in the literature[23]. $S_{ii}$ show linear dependence on $T_e$: $S_{ii} \sim 0.19$ μV K$^{-2}$ × $T_e$; $\rho_{ii}$ is available in the temperature range from 10 K to 500 K in the literature[23]. Outside the temperature range, $\rho_{ii}$ is calculated though linear extrapolation using the experimental data from 400 K to 500 K. With clear dependence of $\alpha_{ii}$ and $\nabla T_e$ on $T_e$, the dependence of $J$ on the excitation intensity can be simulated numerically. The simulation results match experimentally measured excitation intensity dependence as shown in Fig. 2a of the maintext. In the meantime, this data can also be fitted by a power law: $E_{THz} \propto I^{0.66}$ as shown in Supplementary Fig. 5.

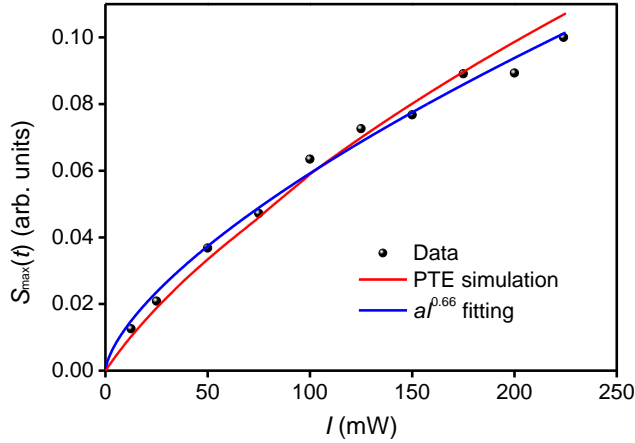

**Supplementary Fig. 5 The laser intensity dependence of the peak field of emitted THz pulse.** The red line is simulated power dependence according to the thermoelectric response calculation, and the blue line is a power law fitting.

We note the above simulation only concerns the thermoelectric response from electrons and the response from holes is not considered. This is because the transport and thermal properties of the holes in $Cd_3As_2$ are not available in the literatures so far. Considering the ambipolar transport nature during the first few picoseconds after the excitation, the contribution of holes to the THz emission cannot be ignore, but it may also follow similar power dependence with electrons according to the experimental results.

When a magnetic field is applied, the Nernst thermoelectric conductivity can be written as $\alpha_{ij} = \sigma_{ii}S_{ij} + \sigma_{ij}S_{jj}$. However, the temperature dependence of the coefficients related to Nernst effect are not available either, so that it is not feasible to simulate numerically. According to Supplementary Eq. 5, $I^{1/2}$ excitation intensity dependence obtained from experiment (as shown in Fig. 3a of maintext) requires the coefficient $\alpha_{ij}$ to have a sublinear dependence on $T_e$ and tends to saturate at high temperature ($T_e^2 \gg T_0^2$, i.e., at high laser intensity).

## Supplementary Note 6.    THz waveform under opposite magnetic field

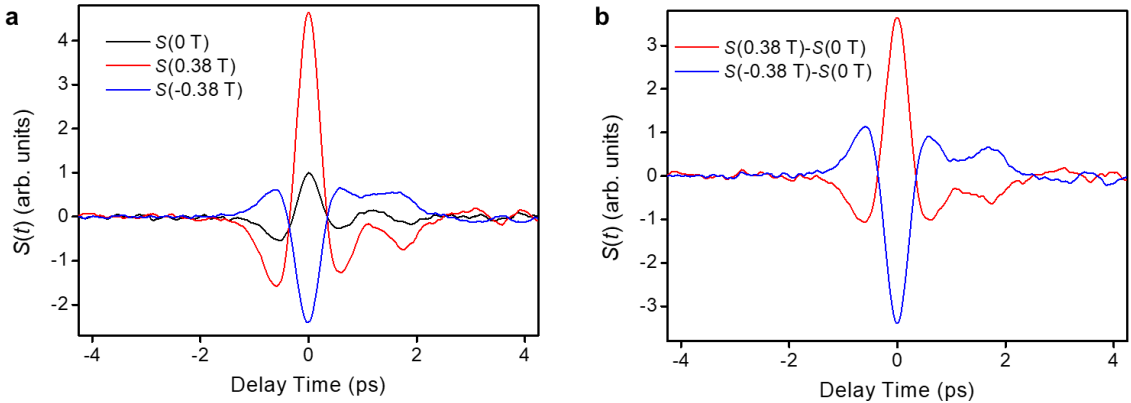

**Supplementary Fig. 6 THz waveform under opposite magnetic field. a** THz emissions from a 130 nm-$Cd_3As_2$/mica sample (which has a relatively large THz signal at 0 T) at $B = 0$ T and $\pm 0.38$ T. **b** The

THz waveforms at $B = \pm 0.38$ T with background signal at $B = 0$ T subtracted. The THz waveforms as result of magnetic field are exactly opposite to each other for $B = \pm 0.38$ T.

**Supplementary Note 7.   Magnetic field dependence of THz waveforms**

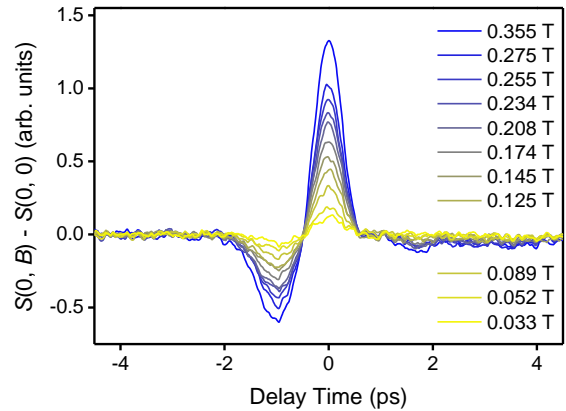

**Supplementary Fig. 7 Magnetic field dependence of THz waveforms.** THz waveforms measured at different in-plane magnetic field intensity with background signal at $B = 0$ T subtracted.

**Supplementary Note 8.   THz emission from $Cd_3As_2$ film**

In this part, we provide experimental evidence that the THz emission is from the $Cd_3As_2$, not from the CdTe buffer layer and the substrate. We measured the THz emission with a 10 nm-CdTe/mica sample as shown in Supplementary Fig. 7. It shows very weak THz emission signal no matter with or without magnetic field, which are about 5 times smaller than the THz emission from the $Cd_3As_2$/CdTe/mica sample without magnetic field. Therefore, this comparison measurement indicates the THz emission is from $Cd_3As_2$ film, not from CdTe buffer layer or mica substrate.

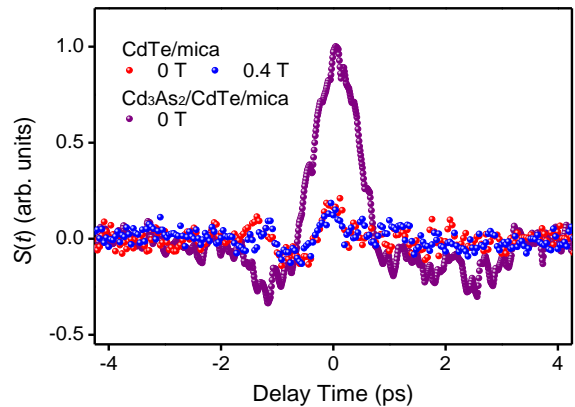

**Supplementary Fig. 8 THz emissions from 10 nm-CdTe/mica at $B = 0$ T and 0.4 T and from $Cd_3As_2$/CdTe/mica at $B = 0$ T.** The peak amplitude of THz emission from $Cd_3As_2$/CdTe/mica is set as 1 in the plot.